\documentclass[12pt]{revtex4}

\usepackage[]{graphicx}
\usepackage{epstopdf}
\usepackage{amsmath}
\usepackage{amsfonts}
\usepackage[latin1]{inputenc} 
\usepackage{amssymb}
\usepackage{makeidx}
\usepackage{tikz}
\usepackage{epstopdf}
\usepackage{subfigure}
\usepackage{multirow,bigdelim}
\newcommand{\beq}{\begin{equation}} \newcommand{\eeq}{\end{equation}}
\newcommand{\bea}{\begin{eqnarray}} \newcommand{\eea}{\end{eqnarray}}
\newcommand{\bear}{\begin{eqnarray*}} \newcommand{\eear}{\end{eqnarray*}}

\begin{document}

	\title{A New Quenched $XY$ model with Nonusual "Exotic" Interactions}
	\author{Anderson A. Ferreira$^1$ }
	
	\affiliation{$^1$ Departamento de  F\'{i}sica, Universidade Federal de S\~ao Paulo, 96010-900, S\~ao Paulo, Brazil}

%
%

\begin{abstract}
Human beings live in a networked world in which information spreads very fast thanks to the advances in technology. In the decision processes or opinion formation there are different ideas of what is collectively good but they tend to go against the self interest of a large amount of agents. Here we show that the associated stochastic operator ($\mathfrak{\widehat{ W^{\aleph}}}_{\rho,\Delta,r}$) proposed in \cite{alef} for describe phenomena, does not belong to a CP (Contact Processes) universality class \cite{hinrichsen2000non}.  However, its mathematical structure corresponds to a new Exotic quantum XY model, but unprecedently,   their  parameters is a "function" of the interaction  between  the local sities $(i-1,i,i+1)$, and  with the impurity present at the same site $i$.
\end{abstract}


\maketitle

\section*{Introduction}

Failure in cooperating can threaten existence itself.  Conflicts and issues such as wars, corruption, loss of liberty and tyrannies, environmental degradation, deforestation, among others pose great problems and are a testament to humanities inability of cooperate in a suitable level. These examples show that we still do not have a complete understanding of the mechanism which drives the collective toward to a common goal and hence to avoid the tragedy of the commons \cite{hardin1968tragedy,ostrom2002drama}. Despite that, altruism, cooperation and moral norms still being improved to outcompete behaviors of free riders, selfishness and immoral \cite{axelrod1981emergence,nowak2006five}.

Living organisms and human beings are characterized by autonomy. However, they tend to be prone to selfishness, a bias that may bring harsh damage to their survival as well as the environment, maybe due to ambitions and potential short-sightedness. The assumption that living organisms are selfish has been accepted by many branches of contemporary science. For example, the inclusive fitness theory, in ecology, in which egoism has biological roots \cite{alexander2017biology}. A similar idea arises in neoclassical economic theory, which hypothesized that all choices, no matter if altruistic or self-destructive, are designed to maximize personal utility \cite{harrison1985egoism}. Thus, decisions are motivate by self-interest.

One of the most interesting issues to be addressed in the context of the present study is morality. The human free will combined with its selfish/altruist nature may create a plethora of different patterns over several types of social systems. How does morality emerge? This question probably does not have a simple and unique answer. Many thinkers from ancient times to today, tries to unravel this phenomenon. As a legacy we have a body of theories that seek to understand and explain the emergence of morality in different societies.  Thomas Hobbes \cite{hobbes2006leviathan,cranston1989jean} was one of the first modern philosopher to offer a naturalist principle to ethics. In his theory, ethics emerge when people understand the necessary conditions to live well. According to Hobbes, these conditions are defined by imposition of equality of rights, by means of an absolute Sovereign, due to the necessity of self preservation and by establishing deals among individuals. Latter on, Rousseau  proposed  that life in community can lead to the loss of individual freedom since the subjects must fulfill a social contract expressed through laws and institutions \cite{rousseau2002social}. Unlike the philosophers who attributed to reason the capacity to conceive morality, Durkheim understood it as a result of a set of social interactions and culture elaborated throughout history \cite{durkheim1973emile,marks1974durkheim}. But these are part of a small selection of seminal works about a theme  hallmarked by an intriguing and challenge scientific problem. This debate continues in different areas such as Psychology \cite{haidt2008social,haidt2008morality}, Political Science, Philosophy, Antropology \cite{ostrom2000collective}, Education, Economics and Ecology \cite{huberman1993evolutionary,santos2006evolutionary}.

To further advance the long discussion on morality or cooperation, we need to understand some specific mechanisms of social interaction in various scenarios with different individual degrees of freedom, effective individual choices, and consider that these choices are influenced locally by peers' s opinions \cite{rowe2002nicomachean,rachels1993elements,brandt1996facts}. Different levels of freedom (free choice), control (supervision) and social dynamics  impact the individual capacity to  fulfillment of the social contract and hence should lead to different degrees of morality or cooperation at the collective level.

It is a well-known fact that a system of interacting linked individuals can work together to reach a collective goal.  Understanding how decentralized actions can lead to these results has been a topic of study in the literature for decades. The focus of our study is the role of a master node, connected to some members of a society, may drive the pursued ideal by collectiveness. The topology formed by a master node connected to a network may represent many situations in social systems: law enforcement and citizens \cite{zaklan2009analysing, orviska2003tax,rose2001trust}; moral and community \cite{rowe2002nicomachean,rachels1993elements,brandt1996facts,haidt2008morality};  beliefs and member of churchs \cite{stark1980networks,shi2016evolution,galam1991towards}; cooperation and egoism \cite{ohtsuki2006simple, beersma2003cooperation}; tax evasion and fiscal country, among others. In all examples, individuals do not share the same goals, due to the incentives in acting against the common good.

We approach this issue using a stochastic quenched disorder model to study the consensus formation \cite{castellano2009statistical,gonzaga2018quenched}. In this model, individuals are autonomous to make decisions based on their own opinions or let decisions be influenced by a local social group or/and by the presence of a norm (master) that reinforces preferential behaviors. The individual decisions are binaries (0 or 1) and the collective decision is the average collective decision.

This model does not belong to a class of nonequilibrium systems \cite{sampaio2011block, medeiros2006domain,stinchcombe2001stochastic}. We found absorbing states phase transitions with respect to three distinct order parameter \cite{dickman1998self,evans2000phase,hinrichsen2000non,de2015generic}. From a statistical mechanics point of view, phase transition in nonequilibrium sytems are studied by fundamental concept as scaling and universality class \cite{lubeck2004universal,lubeck2003universal,odor2004universality,de2005spontaneous}, which may reserve some unexpected results \cite{hexner2015hyperuniformity,van2002universality}.

The remainder of the paper is organized as it follows: in Sec. \ref{sec:model} we define the model and introduce the general notation. In Sec. \ref{sec:sn} we change the notations to construct the new operator. For instance, we use the special parametrization \cite{alef} and the Schutz's prescription \cite{schu} to write a XY model with a new exotic topological interaction. Finally, in Sec. 
\ref{sec:conclusion} contains conclusion and an outlook on future work.

\section{The Model}
\label{sec:model}

\begin{figure}[htb]
	\begin{center}
		\includegraphics[width=0.4\textwidth]{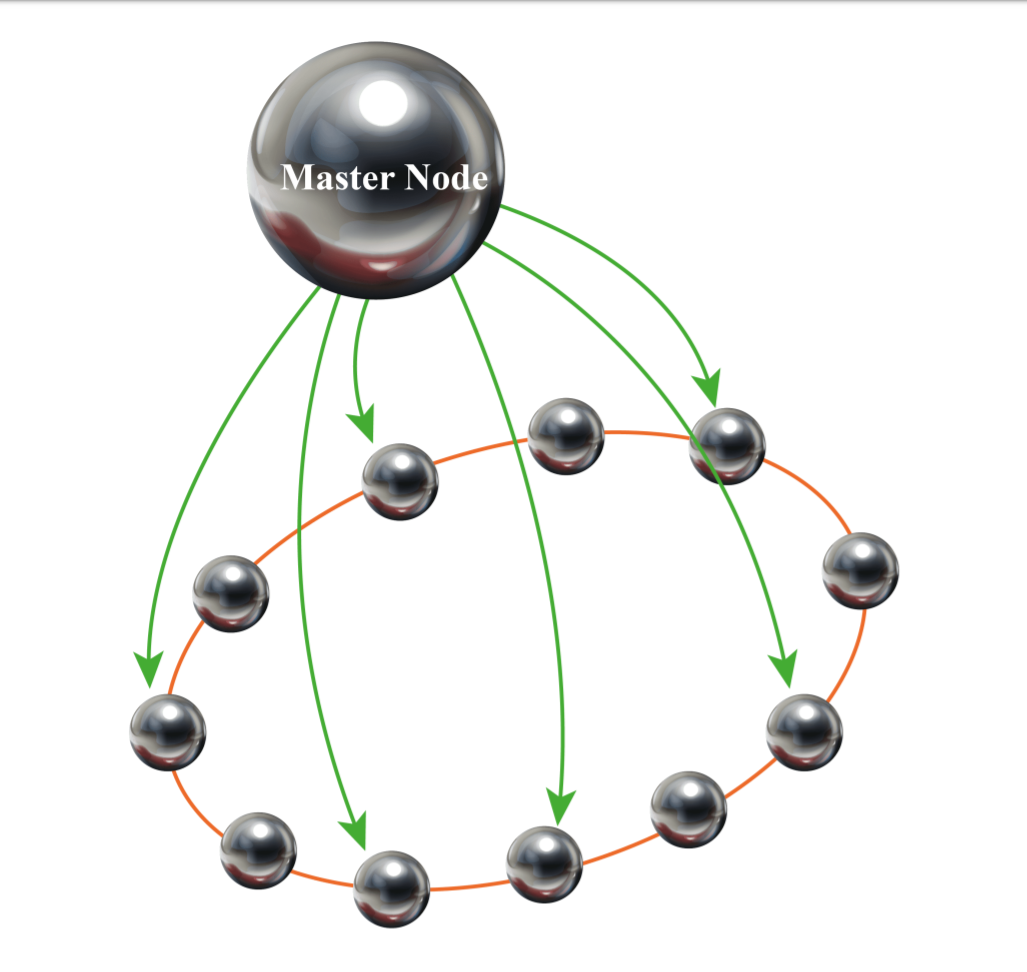}
	\end{center}
	\caption{ Model representation. Small spheres represent interacting particles or agents. Each particle is in the state $s=0,1$ with probability $w_s$.Large sphere respresent the master state, in which the interaction strength with particles is fixed (denoted by $r$) and the links are quenched disorder with density denoted by $\rho$.}
	\label{fig:model}
\end{figure}

In Figure \ref{fig:model} we illustrate our model. It consists of a ring formed by  nodes with periodic boundary conditions. Initially, each node represents particles or agents which may assume two different states $s=0,1$  with probability $w_s$. If , this means that there may be an intrinsic tendency or preference of particles for a determined state $s$. So, in principle, this is a particle property. They interact with first neighbors (on the left and on the right). Moreover, we introduce a master node illustrated by a large sphere on the top of the ring in Figure \ref{fig:model}. The master node connects with particles located in the ring with probability  in the initial time (quenched disorder). The interaction strength between master and connected agents is denoted by $r$. The general configuration of the system is given by $(\varphi_i, \Gamma)$, with $i=1, \dots, L$ representing the individual states and $\Gamma$ representing the existence of a connection between master node and node $i$.

We have two different types of interaction. First of all we identify the interaction between particle $i$ and its neighbors $i-1$ and $i+1$ with the state of particle $i$ dependent on the state of its neighbors $(\varphi_{i-1},\varphi_{i+1})$. In the absence of a master, particle $i$ will align with the majority in the neighborhood, a situation which lead to consensus. If there are differences between neighbor states (frustration), the decision is probabilistic. If the particle $ i $  is in the state $ s $, she/he switches to another  state with probability $ w_s $. When there is a mixed state $(0|1)$,  the probability transition depends on the state  of of the particle $i$.

The second type occurs with presence of the master node $\Gamma_i=1$, which belief or orientation is equal to 1.  The probability of the particles being influenced by the master particle is equal to $ r $. Suppose the particle $i$ is in the state $s = 1$. If most of your neighbors are in the state $s = 0$ then the particle will change to state $s=0$ with probability $ 1-r $. However, if there is frustration between the neighbors $(0 | 1)$ or $(1 | 0)$, then the particle $i$ changes to the state $(s = 0)$ with probability $r_1$. Now suppose that the particle $i$ is in the state $s = 0$. In the case where neighbors are in the state $s=1$, due to peer pressure (majority) and also due to the influence of the master, the particle will change to state $s=1$. There are two conflict situation. When the majority of neighbors are in the state $s=0$, with probability $q$ the particle $i$ will change to state $s=1$ or remain in the same state with probability $p=1-q$. The second situation there is frustration between neighbors. Now, with probability $r_0$ particle $i$ will change to state $s=1$ and stay in the same state with probability $r_1=1-r_0$.

\section{The Second Notation}
\label{sec:sn}
In this point we shall change the notations to contruct the new stochastic operator in the more clean way. We write now the global state of the system by

\begin{eqnarray}
\vec{\mathrm{s}}=(\mathrm{s}_1;\mathrm{s}_2;\ldots;\mathrm{s}_L),
\end{eqnarray}

\noindent{where $\mathrm{s}_i=(c_i,f_i)$.  The variable $c_i$ assume the
	value  $\mathbf{1}$ when the citizen  at the site  $i$ is in the moral state, or the value $\mathbf{0}$ when the citizen is in the immoral state. Besides that, if at site $i$ there is a fiscal (Regulations, Laws, Norms, Contracts, etc) we 
	write $f_i=1$ (YES), otherwise $f_i= 0$ (NO)}.

The dynamic at site $ i $ is not directly influenced by the presence of fiscals in the neighborhood $ i-1 $ and $ i + 1 $. 
At each time step ($\Delta t=1/L$) a site $i$, subject to fiscalization $f_i=\epsilon$, is chosen randomly among $L$ sites of the lattice. The transition probabilitity  of citizen in the state $c_i=c_{out}$ goes to state $c_i=c_{in}$ at the time $t+\Delta t$, given that the states of neighborhood $c_{i-1}=c_{L}$ and $c_{i+1}=c_{R}$ are remained unchanged and the state of fiscals

\begin{eqnarray}
\label{eq:e1}
P(c_i=c_{in};t+\Delta t|c_{i-1}=c_L,c_i=c_{out},c_{i+1}=c_{R},f_i=\epsilon;t):=\mathcal{P}^{c_L\;\;c_R}_{c_{out}\;c_{in}}(\epsilon).
\end{eqnarray}

In Table 1 we show the transition probabilities for situations where an individual located in a certain $ i $ site is not influenced by supervision in the evolution of his state.

\begin{table}[h]
	\centering
	\caption{Probability Transition without fiscalization}
	\label{v0}
	\begin{tabular}{|cc|c|c|c|c|c}
		\cline{3-6}
		\multicolumn{2}{c|}{}&\multicolumn{4}{|c|}{$c_{out} \rightarrow c_{in}$} \\
		\hline
		$c_L$  & $c_R$        & $0 \rightarrow 0 $                     & $0 \rightarrow 1 $                 & $1 \rightarrow 0$                 & $1 \rightarrow 1 $    \\
		\hline
		0      &    0         & 1                & 0                  & 1                  & 0       \\
		\hline
		1      &    1         & 0                & 1                   & 0                   & 1      \\
		\hline
		0      &    1         & \multirow{2}{*}{$1-w_0$} & \multirow{2}{*}{$w_0$} & \multirow{2}{*}{$w_1$} & \multirow{2}{*}{$1-w_1$}\ \\
		1      &   0          &                   &                   &                   &   \\
		\hline
	\end{tabular}
\end{table}

In terms of the notation $(\ref{eq:e1})$ we have

\begin{eqnarray}
\label{eq:e2}
\mathcal{P}^{c_L\;\;c_R}_{c_{out}\;c_{in}}(0)&=&(1-c_{in})(1-c_L)(1-c_R)+c_{in}c_Lc_R+\nonumber\\
&&+\Big{[}(1-w_0)(1-c_{out})(1-c_{in})+w_1(1-c_{in})c_{out}+\nonumber\\
&& +w_0(1-c_{out})c_{in} +   (1-w_1)c_{out}c_{in}\Big{]}|c_L-c_R|.
\end{eqnarray}

While in table 2 we show the transition probabilities for situations where an individual located in a certain $ i $ site is influenced by supervision in the evolution of his state. And in the same way, we have

\begin{table}[h]
	\centering
	\caption{Probability Transition with fiscalization}
	\label{v1}
	\begin{tabular}{|cc|c|c|c|c|c}
		\cline{3-6}
		\multicolumn{2}{c|}{}&\multicolumn{4}{|c|}{$c_{out} \rightarrow c_{in}$} \\
		\hline
		$c_L$  & $c_R$        & $0 \rightarrow 0 $                     & $0 \rightarrow 1 $                 & $1 \rightarrow 0$                 & $1 \rightarrow 1 $    \\
		\hline
		0      &    0         & $1-q$               & $q$                  & $1-r$                  & $r$       \\
		\hline
		1      &    1         & 0                & 1                   & 0                   & 1      \\
		\hline
		0      &    1         & \multirow{2}{*}{$1-r_0$} & \multirow{2}{*}{$r_0$} & \multirow{2}{*}{$r_1$} & \multirow{2}{*}{$1-r_1$} \\
		1      &   0          &                   &                   &                   &   \\
		\hline
	\end{tabular}
\end{table}

\begin{eqnarray}
\label{eq:e3}
&&\mathcal{P}^{c_L\;\;c_R}_{c_{out}\;c_{in}}(1)=
\Big{[}(1-q)(1-c_{out})(1-c_{in})+(1-r)c_{out}(1-c_{in})+\nonumber\\
&&+q(1-c_{out})c_{in}+rc_{out}c_{in}\Big{]}(1-c_L)(1-c_R)+c_{in}c_Lc_R+\nonumber\\
&&+\Big{[}(1-r_0)(1-c_{out})(1-c_{in})
r_1(1-c_{in})c_{out}+
r_0(1-c_{out})c_{in} +\nonumber\\ 
&&+(1-r_1)c_{out}c_{in}\Big{]}|c_L-c_R|.
\end{eqnarray}

\subsubsection*{Parameterization}

Let us choose some constraints to the parameters $p, q, r_0, r_1, w_0$ and $w_1$ in terms of $r$ ('' influence of master node'') and $\Delta=w_0-w_1$, which is the intrinsic state tendency of agents, and $\rho$. For simplicity, we take $q=r$.  Since $w_1=1-w_0$ we may write

\begin{equation}
\Delta=w_0-w_1 =1-2w_1.
\end{equation}

The parameter $\Delta$ measures the natural nature of an element or particle be in the state $s=0 (1)$ when $\Delta>0 (<0)$ in the absence of any interaction or influence.
If $0<w_1<\frac{1}{2} $  the individuals, in average, will behave against the norm or the common good. In this case, $\Delta>0$ which means that the system has a tendency to be opposite to the master (selfish or immoral). We are interested in studying how such a system undergoes to a phase dominated by the main orientation (cooperative or exclusively moral), so we will vary the parameter in the interval $0<\Delta<1$.


The probabilities $r_0$ and $r_1$ should be parameterized so that when the master's influence is null ($r=0$) we have $r_0=w_0$ and $r_1=w_1$. Otherwise, when $r=1$  we  should have necessarily $r_0=1$ and $r_1=0$. The simplest way is through a linear parameterization

\begin{eqnarray}
&&r_0=r+(1-r)(\frac{1-\Delta}{2}),\\
&&r_1=(1-r)(\frac{1+\Delta}{2}).
\end{eqnarray}

\noindent The parametrized version of the model has only three free parameters:

\begin{equation}
0\leq r \leq 1,\;\;\;\;0\leq \Delta \leq 1\;\;\;\mbox{and}\;\;\;\;\; 0\leq\rho\leq 1.
\end{equation}

\noindent{Taking into account such parametrizations, the $(\ref{eq:e2})$ e $(\ref{eq:e3})$ equations can be grouped in a more compact form, that is,}

\begin{eqnarray}
\label{eq:special}
&&\mathcal{P}^{c_L\;\;c_R}_{\;\;\;\;c_{in}\;}(f_i)=
[(1-rf_i)(1-c_{in})+rf_ic_{in}](1-c_R)(1-c_L)+c_{in}c_Lc_R+\nonumber\\
&+&{\frac{(1-rf_i)(1+\Delta-2\Delta c_{in})+2rf_i}{2}}|c_L-c_R|.
\end{eqnarray}

\section{Quantum chain}
\label{sec:qc}


The time evolution of the probability $|P(t)\rangle$ correspondent to stochastic state $|\beta\rangle$ at time $t$ is governed by markovian transfer operator $W$ \cite{legeto}.
Writing the master equation in its continuous-time differential form, we have

\begin{eqnarray}
\label{eq:master1}
\partial t \left|P({\sigma},t)\right\rangle=\sum_{\beta}w_{(\beta\rightarrow\sigma)}\left|P({\beta},t)\right\rangle-
w_{(\sigma\rightarrow\beta)}\left|P({\sigma},t)\right\rangle ,
\end{eqnarray}

\noindent{where $\sigma,\beta$ represent two distinct lattice configuration. 
	Rewriting  the equation ($\ref{eq:master1}$) in its vector form \cite{schu}
	
	\begin{eqnarray}
	\label{eq:master2}
	\partial_t \left|P\right\rangle=-W\left|P\right\rangle,
	\end{eqnarray}
	
	\noindent{where $W$ is a matrix operator, responsible for connecting differents configurations of the vector space. 
		It is also important to mention that, in general, this operator is not Hermitian, i.e., it has complex eigenvalues. 
		These eingenvalues correspond to the oscillations in the model (imaginary part), while the 
		exponential decay is contained in the real part.}

	\noindent{In an orthonormal basis we have $\left\langle \sigma^n\right|\left|\beta^n\right\rangle=\delta_{\sigma_1,\beta_1}\delta_{\sigma_2,\beta_2}\cdots\delta_{\sigma_n,\beta_n}$. This suggests that we can
		write $\left|P\right\rangle$ as}
	
	\begin{eqnarray}
	\left|P\right\rangle=\sum_{\beta}P(\beta,t)\left|\beta\right\rangle.
	\end{eqnarray}
	
	\noindent{If we denote the initial probability of the system by $\left|P_o\right\rangle=\sum_{\beta}P_o(\beta)\left|\beta\right\rangle,$ 
		the formal solution of the problem can be written as }
	
	\begin{eqnarray}
	\left|P\right\rangle=\left|P_o\right\rangle=e^{-Wt}\left|P_o\right\rangle.
	\end{eqnarray}
	
	Due to conservation of probability, we have $\left\langle 0\right|W=0$, where $\left\langle 0\right|=\sum_{\beta}\left\langle \beta\right|$. Thus any
	observable can be calculated as follows
	
	\begin{eqnarray}
	&&<X>_t=\sum_{\beta}X(\beta)P(\beta,t)\left|\beta\right\rangle=\nonumber\\
	&&\left\langle 0\left| X\right|P \right\rangle=\left\langle 0\left| Xe^{-Wt}\right|P_o \right\rangle.
	\end{eqnarray}

	Here, we always can choose  a physical intuitive "Canonical Base" 
	$\mathcal{B}$$= \{\left|\beta\right\rangle_1,\left|\beta\right\rangle_2\ldots,\left|\beta\right\rangle_L\}$ to construct the Hilbert  Physical Space, i.e;

	\begin{eqnarray}
	\left|\beta\right\rangle=\left|\beta\right\rangle_1\bigotimes\left|\beta\right\rangle_2\bigotimes \ldots \bigotimes \left|\beta\right\rangle_L, 
	\end{eqnarray}
	
	\noindent{where $\left|\beta\right\rangle_i=\left|\beta_I\right\rangle_i\otimes\left|\beta_F\right\rangle_i$. The letter $I$ represents an individual and the letter $F$ represents a fiscal. The vector $\left|\beta_I\right\rangle_i$ can be takes on
		the number  $1$ when the individual at the site  $i$ is in the moral state and $0$ when this is in immoral state. If the site $i$ has a fiscal we will represent this sitituation writting $\left|\beta_F\right\rangle_i=1$, otherwise $\left|\beta_F\right\rangle_i=0$.}

	We now introduce the new stochastic operator related with this model. For instance, we will design the operator $W$ like

	\begin{eqnarray}
	\mathfrak{\widehat{ W^{\aleph}}}_{\rho,\Delta,r}=\sum_{k=2}^{L-1}\mathfrak{\widehat{W^{\aleph}}}_k,
	\end{eqnarray}

	\noindent{where $\mathfrak	{\widehat{W^{\aleph}}}_k$ connects two differents states in the assyncronous dynamics, in others words, the matrix elements can be write as}

	\begin{eqnarray}
	\mathfrak{\widehat{W^{\aleph}}}_k=\Large{\widehat{\mbox{{\bf{1}}}}}\bigotimes\cdots\bigotimes\widehat{\Omega^{\aleph}_k}\bigotimes\cdots\bigotimes \Large{\widehat{\mbox{\bf{1}}}}.
	\end{eqnarray}
	
	The operators ${\widehat{\Omega^{\aleph}}}_k$ act  in the state   $\left|\beta\right\rangle_k$. Assuming periodic boundary conditions $( \left|\beta\right\rangle_0\equiv  \left|\beta\right\rangle_L$ and $( \left|\beta\right\rangle_{L+1}\equiv  \left|\beta\right\rangle_1$), we can separate the element $\left\langle \alpha\right|\mathfrak{\widehat{W^{\aleph}}}_k \left|\beta\right\rangle$ in two contribution, i.e, a contribution $\left\langle \alpha\right|\mathfrak{\widehat{W^{\aleph}}}_k \left|\beta\right\rangle_{NF}$ without fiscal at the site $k$ and a another contribuiton $<\alpha|\mathfrak{\widehat{W^{\aleph}}}_k \left|\beta\right\rangle_{F}$  with a fiscal at the site $k$. The general matrix element can be write as 
	
	\begin{eqnarray}
		\left\langle \alpha\right|\widehat{\Omega_k}\left|\beta\right\rangle=\left\langle \alpha\right|\widehat{\Psi_k^{NF}}\otimes {\small{\widehat{\mbox{{\bf{1}}}}}}\left|\beta\right\rangle+\left\langle \alpha\right|\widehat{\Phi_k^{F}}\left|\beta\right\rangle
	\end{eqnarray}

	If we use the Schutz's prescription \cite{schu} to construct the stochastic operator in terms of Pauli's matrices, then the operator  $\mathit{\widehat{W^{\aleph}_k}}$ assumes a more elgant form

	\begin{eqnarray}
	&&\mathit{\widehat{W^{\aleph}_k}}=\sum_{\mu,\nu,\gamma=\pm}J_{\mbox{off}}[\hat{v}_{\hat{b_k}}]\widetilde{a^{\mu}_k}
	\widetilde{ a^{\nu}_{k-1}}\widetilde{ a^{\gamma}_{k+11}} +
	J_{\mbox{on}}[r\hat{n}_{\hat{b_k}}]{a^{\mu}_k}
	\widetilde{ a^{\nu}_{k-1}}\widetilde{ a^{\gamma}_{k+11}} +
	{a^{\mu}_k}
	{ a^{\nu}_{k-1}}{ a^{\gamma}_{k+11}}\mbox{\Huge +}\nonumber\\
	&&\mbox{\Huge +}\mbox{\Huge [}f(\Delta)J_{\mbox{off}}[\hat{v}_{\hat{b_k}}]+g(\Delta)J_{\mbox{off}}[\hat{v}_{\hat{b_k}}]a^{\mu}_k+J_{\mbox{on}}[r\hat{n}_{\hat{b_k}}]\mbox{\Huge ]}\mbox{\big ( }a^{\nu}_{k-1}-a^{\gamma}_{k+1}\mbox{\big ) },
	\end{eqnarray}
	
	\noindent{ where}
	
	\begin{eqnarray}
	&&\widetilde{ a^{\nu}_{k}}=1-a^{\nu}_{k},\\
	&&J_{\mbox{on}}[r\hat{n}_{\hat{b_k}}]=r\hat{n}_{\hat{b_k}},\\
	&&J_{\mbox{off}}[\hat{v}_{\hat{b_k}}]=\hat{v}_{\hat{b_k}}=1-r\hat{n}_{\hat{b_k}},\\
	&&f(\Delta)=\frac{1+\Delta}{2}\;\;\;\; \mbox{and}\\
	&&g(\Delta)=-\Delta,\;\;\;\;\;\;\mbox{with}\;\;\;\; [b^{+},b^{-}]_k=0.
	\end{eqnarray}

	Here the operators $a^{+},a^{-}$ act in the subspace $|\beta_I>$ and the operators $b^{+},b^{-}$ act in the subspace $|\beta_F>$.
	However, the most beautiful interpretation is about means of $J_{\mbox{on}}$ and $J_{\mbox{off}}$. We can roughly look at couplings as a kind of function of the "quenched impurity interaction". In the other words, this simple model revels  a new cathegory of interactions in Statistical Mechanics, i.e; the ON-OFF Quenched Interaction.
	
	In addition, if we just "preserve" the mathematical structure of this operator and choose  an appropriate distributions	$f(\Delta)$ and $g(\Delta)$ with $\Delta$ and $r \in \Re$, then it is possible, for  $a^{\pm}_k=\frac{\sigma^x_k+\sigma^x_k}{2}$, to map the operator $\mathfrak{\widehat{W^{\aleph}}}$ onto  a new class of ${\widehat{H_{XY}}}$ models, i.e; the ${\widehat{H_{XY}^{\aleph}}}$ Queched Model with Special Topological Interactions. 
	
	\section{Conclusion}
	\label{sec:conclusion}
	In the present work we proposed a stochastic quenched disorder model to investigate the power of a master node over a system formed by $L$ elements disposed in a ring network with first neighbor interaction. Due the map between the Master Equation \cite{kanpen} and the Schr\"odinger equation \cite{legeto,alef0} it is possible connect a stochastic one-dimensional model in a quantum chain model. 
	Through the Schutz's protocol \cite{schu},  we got map the stochastic operator in a new  XY quenched model wtith special exotic (on-off) interactions. 	 	All the questions addressed go beyond the parametrization studied here.

	Although the rules of interaction are simple, we uncover a rich scenario of collective behaviors. The major evidence is given by the phase diagram presented in \cite{alef}. The model analyzed here shows the existence of critical values in several parameters. We try to illustrate the volume of the phase space which the coordinates are the control parameter $ \Delta, ~r$ and $\rho$. In the inner part of this volume the order parameter reach its maximum value $M=1$. The shape in this figure is just illustrative. What calls our attention is the properties of the surface of this volume: it separates the synchronized phase where every elements enter in the absorbing state $s=1$ and the phase where there is a mixture $0<M<1$. This idea is corroborated by Figure \cite{alef}. We fixed a plan by choosing specific values of $\Delta$. After, we varied $\rho$ and $r$ and we found a critical line splitting two phases. This imply the existence of a critical surface in the 3D phase diagram.
	
	The critical exponents  $\Lambda_c$  along the manifold surface likely are non universals since they may exhibit a continuous dependence of the exponents with the critical control parameters  $\Lambda_c (\Delta,r,\rho)$. This phenomena is represented by small lines leaving the critical surface \cite{alef} just to give some ideal of a richness of the phase transition occurring in this system.


	\begin{figure}[htb]
		\begin{center}
			\includegraphics[height=5cm, width=8cm]{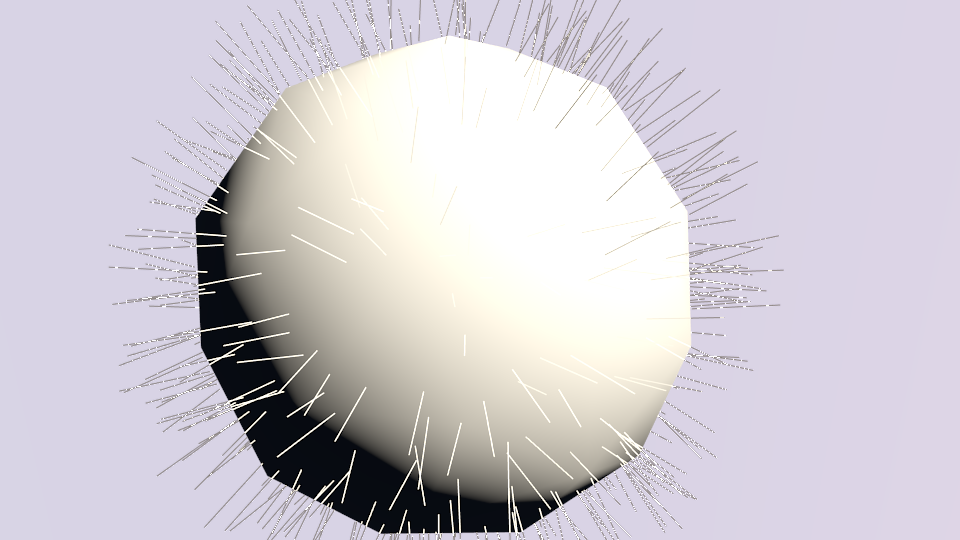}
		\end{center}
		\caption{Illustrative view of a critical surface. The lines leaving the critical surface illustrate dependence of the critical exponent with parameters $\Delta$, $1-\rho$ and $r$.}
		\label{fig:manifold}
	\end{figure}

	\section{Acknowledgments}
	
	The author is partially supported by FAPESP grants. 

	\bibliography{masternodenetwork}
\end{document}